\documentclass[pra,twocolumn,showpacs,a4paper]{revtex4}
\usepackage{bm,graphicx,amsmath,amsfonts,amssymb}

\newcommand{\op}[1]{\boldsymbol{\mathit{#1}}}
\newcommand{\I}{\op{I}}
\newcommand{\Y}{\op{Y}}
\newcommand{\Z}{\op{Z}}
\newcommand{\X}{\op{X}}
\newcommand{\U}{\op{U}}
\newcommand{\ZZ}{\mathbb{Z}}
\newcommand{\XX}{\mathbb{X}}
\newcommand{\ket}[1]{\left|#1\right\rangle}

\begin{document}

\title{Experimental Test of Two-way Quantum Key Distribution
\\in Presence of Controlled Noise}

\author{Alessandro Cer{\` e}}
    %\email{alessandro.cere@unicam.it}
\affiliation{Dipartimento di Fisica, Universit\`a di Camerino,
I-62032 Camerino, Italy}
\author{Marco Lucamarini}
    %\email{marco.lucamarini@unicam.it}
\affiliation{Dipartimento di Fisica, Universit\`a di Camerino,
I-62032 Camerino, Italy}
\author{Giovanni Di Giuseppe}
\affiliation{Dipartimento di Fisica, Universit\`a di Camerino,
I-62032 Camerino, Italy}
\author{Paolo Tombesi}
\affiliation{Dipartimento di Fisica, Universit\`a di Camerino,
I-62032 Camerino, Italy}

\date{\today}

\begin{abstract}
We describe the experimental test of a quantum key distribution
performed with a two-way protocol without using entanglement. An
individual incoherent eavesdropping is simulated and induces a
variable amount of noise on the communication channel. This allows
a direct verification of the agreement between theory and
practice.
\end{abstract}


\maketitle
\begin{figure}%[!ht]
\includegraphics[width=.45\textwidth]{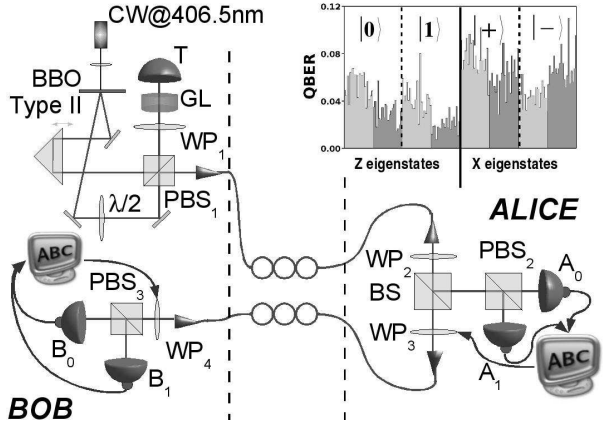}
\caption{\label{fig:setuptomo}Sketch of the experimental setup.
Inset: the distribution of QBER for different sets of preparation
by Bob and operation by Alice. Bob's preparation is reported on
the overlay, Alice's encoding is represented by the area colour:
lighter gray for $\I$ (logical `0'), darker gray for $i\Y$
(logical `1').}
\end{figure}

%%%% Introduction %%%%%
One of the most attractive application of quantum mechanics is the
quantum key distribution (QKD), which allows for the secret
sharing of correlated random data between two (or more) users,
traditionally called Alice and Bob. Since the seminal works by
Bennett and Brassard~\cite{Bennett84} and Ekert~\cite{Ekert91},
QKD developed into a promising field of research for near-future
technology (\cite{addref1}; for a review see~\cite{Gisin02}), and
both theoretical~\cite{addref2} and experimental~\cite{addref3}
work has been done in order to prove its security and feasibility.
A feature of QKD is that none of the users knows in advance the
final form of the generated key. On the contrary the secure
transmission of a predetermined key has been recently investigated
through schemes that exploit~\cite{Long02,EntPP,Bos02} or do not
exploit~\cite{NotEntPP,Lucamarini05} entanglement, and that are
usually cited as \textit{deterministic}, with reference to Bob's
\textit{in principle} possibility of knowing with certainty the
information encoded by Alice~\cite{foot1}.
\\
\indent In this Letter we present the experimental test of a QKD
realized with the two-way protocol described in
Ref.~\cite{Lucamarini05} and termed LM05. We simulate the noise
related to a class of attacks by Eve, and by accordingly varying
it we measure all the
quantities relevant to an effective transmission of information.\\
%%%%% The Protocol %%%%%%
\noindent \emph{The Protocol}-- In LM05 Bob prepares a qubit in
one of the four states $|0\rangle $, $| 1\rangle $ (the Pauli $\Z$
eigenstates), $| +\rangle $, $| -\rangle $ (Pauli $\X$
eigenstates), and sends it to his counterpart Alice. With
probability $c$ Alice uses the qubit to test the channel noise
(\textit{control mode}, CM) or, with probability $1-c$, she uses
it to encode a bit of information (\textit{encoding mode}, EM).
The CM consists in a projective measurement of the qubit along a
basis randomly chosen between $\Z$ and $\X$, followed by the
preparation of a new qubit in the same state as the outcome of the
measurement. The EM is the modification of the qubit state
according to one of the following transformations: the identity
operation $\I$, which leaves the qubit unchanged and encodes the
logical `0', or $i\Y\equiv \Z\X$, which flips the qubit and
encodes the logical `1'. Alice can now send the qubit back to Bob
who measures it in the same basis he prepared it; in case of an EM
run this feature allows Bob to deterministically infer Alice's
operation. After the whole transmission Alice declares the CM  and
the EM runs. Comparing the data collected during the CM the users
estimate the Quantum Bit Error Rates (QBERs) on the forward and
backward channels; we call these two `partial' QBERs respectively
$q_{1}$ and $q_{2}$. Comparing a part of the data collected during
the EM the users estimate also a third, `total', QBER $Q_{AB}$.
This quantity is not necessary for the security of the protocol
but proves useful for estimating the mutual information between
Alice and Bob. As usual for a QKD the exchange of the raw key is
then followed by the procedures of error correction~\cite{ECC} and
privacy amplification~\cite{PA}.\\
%%%%%% The setup %%%%%%%
\indent The communication is realized exploiting linear
polarization states of near infrared photons.
In Fig.\ref{fig:setuptomo} is reported a sketch of the
experimental setup. The photons are generated by a type II
down-conversion process: a UV pump beam from a diode laser
(%Coherent Compass 405-25,
$\lambda=406.5$nm, power 25mW) impinges on a $1.5$mm thick BBO
crystal cut at  an angle of $43^{\circ}$. We select the
intersection of the ordinary propagating light cone and the
extraordinary one at an angular aperture of the two beams around
$\sim3.5^{\circ}$~\cite{Rubin96}. The wide spectral width of the
pump beam ($\Delta\lambda=0.9$nm) affects the entanglement of the
two-photon, as in the case of pulsed pump~\cite{Grice97}; to
obtain a totally symmetrical spectrum we used an interferometric
technique~\cite{Kim03b}. The two output modes of $\text{PBS}_1$
are launched into single mode fibers~\cite{Kurtsiefer01} at
$810$nm through a pair of  interference filters centered at
$810$nm and a bandwidth of $40$nm, which are used to reduce the
background light. The fibers terminate onto the sensible area of
two APD modules with quantum efficiency $\sim70\%$ at $810$nm. The
coincidence rate is around $1000$cps for single count rate of
around $12000$cps. The polarization state after the $\text{PBS}_1$
can be expressed as $ \ket{\psi}=( \ket{00}-\ket{11})/\sqrt{2}$.
Two Glenn-Laser polarizer (GL) were inserted after the
$\text{PBS}_1$ to verify the quality of the polarization
entanglement of the state. The state purity has been tested by a
tomographic reconstruction~\cite{James01}. We measured an
entanglement of formation of $0.989\pm0.005$~\cite{Wootters98} and
a violation of the CHSH Bell inequality of over 98 standard
deviations~\cite{Weihs98}.
Bob prepares the qubits measuring one of the output mode of the
interferometer with a $\lambda/2$ waveplate ($\text{WP}_1$), a GL
and detector T. The detection of this photon projects the other
one on the desired polarization state and serves also as trigger
for the whole communication system. We note that entanglement is
not necessary for the protocol itself. This specific preparation
procedure was chosen because it can be easily extended for a true
random passive choice of the initial state using only linear
optics components. The prepared photon is then launched into a
$5$m long single mode fiber at $810$nm towards Alice.
\\
\indent Alice passively switches between CM and EM via a 50/50 BS.
\emph{Control mode}: when Alice's measuring basis is the same as
Bob's the coincidence clicks between the two detectors A$_0$ and
A$_1$ and the trigger detector T permit to estimate $q_1$ through
the usual formula: $q_1=C_{err}/(C_{0}+C_{1})$, where $C_{0,1}$ is
the rate of coincidence and $C_{err}=C_{0,1}$ depending on Bob's
state preparation. To complete the control mode, Alice injects an
attenuated light pulse of definite polarization with a wavelength
of 810nm in the BS. This pulse is generated by a pulsed diode
laser (not shown in figure) with a repetition rate of 80MHz, pulse
width 88ps FWHM, attenuated to an average number of photons per
pulse of $\mu=(1.20\pm0.05)\times10^{-3}$. \emph{Encoding Mode}:
for encoding the message is necessary to realize the $\I$ and the
$i\Y$ Pauli operators. A couple of $\lambda/2$ waveplates
($\text{WP}_{2,3}$) allows to span all the equator of the Bloch
sphere (an eventual phase has no importance for the rest of the
protocol). As last step, the photon travels back to Bob through a
different fiber, $5$m long too. The photons sent by Alice are
eventually polarization-analyzed at Bob's side by $\text{PBS}_3$
and a $\lambda/2$ waveplate ($\text{WP}_4)$ set so that the
photons are measured in the same basis as they were prepared. The
photons are collected after the $\text{PBS}_3$ into two multimode
fibers and then detected by two APD modules, B$_0$ and B$_1$.
During CM runs a measure of $q_2$ is obtained in the same way as
$q_1$ with the difference that the photons come from the pulsed
laser that also supplies  a trigger signal.
Coincidence counts between B$_0$ or B$_1$ and the trigger T can be
associated to logical values `0' or `1' corresponding to Alice
encoding in the EM runs.
In our experimental tests we used all the coincidence counts to
estimate $Q_{AB}$. In the inset of Fig.\ref{fig:setuptomo} is
reported a typical communication test. It consists in a direct
measurement of $Q_{AB}$ for different state preparations performed
by Bob and different encodings by Alice. All the eight
configurations of interest are reported. The best value we
obtained for $Q_{AB}$ is $(4.05\pm0.22)\times10^{-2}$. Before
every test the fibers were aligned using two polarization control
pads, one for each fiber. The pads are set so that any
polarization input state exits almost unchanged. The usual
fidelity for the polarization state after the alignment is
$\sim96\%$. The fibers proved to remain stable for quite long
periods ($\sim4$h), enough for several runs after the
alignment~\cite{Poppe04}.
\\
%%%%% The Eavesdropping %%%%%%%%
\emph{Eavesdropping}-- The CM of LM05 comprises the same security
test of BB84~\cite{Bennett84}, repeated twice. This gives to the
two-way LM05 at least the same security level of the one-way BB84.
Nevertheless there are indications that the security threshold of
two-way schemes can overcome that of one-way schemes. In
particular LM05 results secure against individual incoherent
attack (IIA) regardless of the noise introduced on the channel by
an eavesdropper (Eve)~\cite{Lucamarini05}; on the contrary BB84
results secure against individual attacks only if the noise
threshold is lower than $\sim$15\% (\cite{Fuc97},~\cite{Gisin02}
Sec.VI.E). In the optimal IIA Eve prepares two sets of ancillae
$\varepsilon,\eta$ and makes them interact with the qubit: the
$\varepsilon$'s on the forward path and the $\eta$'s on the
backward one, after Alice's encoding stage. By proper measure of
her two sets of ancillae, Eve can gain information about the key.
There are two \textit{mutually exclusive} interactions that
minimize Eve's noise on the channel while maximizing her
gain~\cite{Lucamarini05}:
\\
$\ZZ$-attack: \vspace{-1.5mm}
\begin{equation}\label{Z}
\begin{alignedat}{3}
\ket{0}\ket{\varepsilon} &\rightarrow \ket{0}\ket{\varepsilon_{0}^{\ZZ}} &\;\qquad&
 \ket{+}\ket{\varepsilon}&\rightarrow \ket{+}\ket{\varepsilon_{+}^{\ZZ}}+\ket{-}\ket{\varepsilon_{-}^{\ZZ}}\\
\ket{1}\ket{\varepsilon} &\rightarrow \ket{1}\ket{\varepsilon_{1}^{\ZZ}} &\;\qquad&
 \ket{-}\ket{\varepsilon}&\rightarrow \ket{+}\ket{\varepsilon_{-}^{\ZZ}}+\ket{-}\ket{\varepsilon_{+}^{\ZZ}}
\end{alignedat}
\end{equation}
$\XX$-attack: \vspace{-1.5mm}
\begin{equation}\label{X}
\begin{alignedat}{3}
\ket{+}\ket{\varepsilon}&\rightarrow \ket{+}\ket{\varepsilon_{+}^{\XX}} &\;\qquad&
 \ket{0}\ket{\varepsilon} &\rightarrow \ket{0}\ket{\varepsilon_{0}^{\XX}}+\ket{1}\ket{\varepsilon_{1}^{\XX}}\\
\ket{-}\ket{\varepsilon}&\rightarrow \ket{-}\ket{\varepsilon_{-}^{\XX}} &\;\qquad&
 \ket{1}\ket{\varepsilon} &\rightarrow \ket{0}\ket{\varepsilon_{1}^{\XX}}+\ket{1}\ket{\varepsilon_{0}^{\XX}}
\end{alignedat}
\end{equation}
where we have introduced the states $\vert
\varepsilon_{0,1}^{\ZZ}\rangle$, $\vert
\varepsilon_{+,-}^{\XX}\rangle$, $\vert \varepsilon_{+,-}
^{\ZZ}\rangle =[\vert \varepsilon_{0}^{\ZZ}\rangle \pm \vert
\varepsilon_{1}^{\ZZ}\rangle]/2$ and $\vert
\varepsilon_{0,1}^{\XX}\rangle =[\vert
\varepsilon_{+}^{\XX}\rangle \pm\vert
\varepsilon_{-}^{\XX}\rangle]/2$. The $\ZZ$ and $\XX$  indicate
the basis whose eigenstates remain unchanged under Eve's action.
The states $\vert \varepsilon_{0,1}^{\ZZ}\rangle$ and $\vert
\varepsilon_{+,-}^{\XX}\rangle$ are normalized and non orthogonal:
$\langle\varepsilon_{0,1}^{\ZZ}|\varepsilon_{1,0}^{\ZZ}\rangle
=\cos \phi_{\varepsilon}^{\ZZ}$ and $\langle
\varepsilon_{+,-}^{\XX}|\varepsilon_{-,+}^{\XX}\rangle
=\cos\phi_{\varepsilon}^{\XX}$, with
$\phi_{\varepsilon}^{\ZZ,\XX}\in[ 0,\pi/2]$. Analogous expressions
hold for backward propagation with $\eta$'s ancillae in the place
of $\epsilon$'s. If Eve chooses the $\ZZ$-attack, Eqs.(\ref{Z}),
she introduces no noise on the channel when Bob prepares the
eigenstates of the $Z$-basis but creates disturbance in the
conjugate basis $X$; the same argument applies for the
$\XX$-attack, Eqs.(\ref{X}). The absence of a public basis
revelation in the LM05 protocol prevents Eve from always choosing
the best attack strategy between the $\ZZ$ and the $\XX$-attack.
In the frame of the IIA attack an explicit functional relation
among the three QBERs, $q_{1}$, $q_{2}$ and $Q_{AB}$ can be found.
Let us consider the expression of the forward QBER for the $Z$ and
$X$-state preparation, $q_{1z}$  and $q_{1x}$ respectively, as
functions of the angles $\phi_{\varepsilon}^{\ZZ}$ and
$\phi_{\varepsilon}^{\XX}$ chosen by Eve on the forward path:
\begin{equation}\label{q1q2}
    \begin{aligned}
    &q_{1z}
        = (1-\cos\phi_{\varepsilon}^{\XX})/2 & \quad
&    &q_{1x}
        = (1-\cos\phi_{\varepsilon}^{\ZZ})/2
\end{aligned}
\end{equation}
Through these relations Alice and Bob can guess Eve's angles
$\phi_{\varepsilon}^{\XX},\phi_{\varepsilon}^{\ZZ}$ from the
measured quantities $q_{1z},q_{1x}$, respectively. Analogous
results (with angles $\phi_{\eta}^{\XX}$ and $\phi_{\eta}^{\ZZ}$)
hold for the partial QBERs $q_{2z}$ and $q_{2x}$ of the backward
channel. The expression of the third QBER $Q_{AB}$ can be derived
as the average probability that Alice and Bob find an error on the
total two-way channel in the EM:
\begin{equation}\label{Eq:QAB_Z}
  Q_{ABi} = q_{1i}+q_{2i}-2q_{1i}\,q_{2i}
        \qquad\quad i=(x,z)
\end{equation}
This relation is suitable for direct verification since the QBERs
on the left side and those on the right side are measured through
independent processes, i.e.~respectively during EM and CM.\\
\indent To simulate the presence of an eavesdropper we must
control the noise on the channels to generate the same effect
caused by Eve's action described in Eqs.\eqref{Z} and~\eqref{X}.
Consider the following unitary transformation:
\begin{equation}\label{eq:Uz}
\U_{\phi_{\varepsilon}^{\ZZ}}^{\ZZ}
        =\cos\phi_{\varepsilon}^{\ZZ}\,\I
           +i \sin\phi_{\varepsilon}^{\ZZ}\,\Z\,,
\end{equation}
where $\phi_{\varepsilon}^{\ZZ}$ is the same angle defined for the
Eve's $\ZZ$-attack. Following the action of
$\U_{\phi_{\varepsilon}^{\ZZ}/2}^{\ZZ}$ on the input states of the
$X$-basis, we find that they are flipped with probability
$\sin^{2}\left(  \phi_{\varepsilon}^{z}/2\right) =\left( 1-\cos
\phi_{\varepsilon}^{z}\right)/2$, equal to the expression of
$q_{1x}^{\ZZ}$ in Eq.\eqref{q1q2}. Therefore the unitary
transformation $\U_{\phi_{\varepsilon}^{z}/2}^{\ZZ}$ determines on
the forward channel the same effect as Eve's attack. An analogous
result is true for the backward path. To evaluate the total QBER
$Q_{AB}$, we must consider the transformations on the $X$-states
for the forward and backward channel,
$\U_{\phi_{\varepsilon}^{\ZZ}/2}^{\ZZ}$ and
$\U_{\phi_{\eta}^{\ZZ}/2}^{\ZZ}$, and both Alice's encoding
operations, $\I$ and $i\Y$. In this way we find the two
expression: $Q_{ABx}^{\ZZ}(\I) =
\sin^{2}(\phi_{\varepsilon}^{\ZZ}/2 +\phi_{\eta}^{\ZZ}/2)$ and
$Q_{ABx}^{\ZZ}(i\Y) =\sin^{2}(\phi_{\varepsilon}^{\ZZ}/2
-\phi_{\eta}^{\ZZ}/2)$. This quantities depend on Alice's
transformation, but if we take the average between them we find
the expression $\overline{Q}_{ABx}^{\ZZ} = (1 -
\cos\phi_{\varepsilon}^{\ZZ}\cos\phi_{\eta}^{\ZZ})/2$, exactly
equal to Eq.\eqref{Eq:QAB_Z} after expressing the partial QBERs
through Eqs.~\eqref{q1q2}.
\\
\indent The simulated eavesdropping described by Eq.\eqref{eq:Uz}
is realized via the two polarization controlling pads on the
fibers connecting Alice and Bob. The $\ZZ$-attack is achieved
aligning the pads so that the $Z$-states remain almost undisturbed
during the propagation. The $\XX$-attack is obtained in a similar
fashion. The presence of undesired contributions due to $\Y,\X$
operators during the simulated $\ZZ$-attack are taken into account
through a parameter $\Delta$, which quantifies the distance from a
perfect realization of the unitary transformation of
Eq.\eqref{eq:Uz}. Once the two fibers are aligned, we measured the
three QBERs $q_1$, $q_2$ and $Q_{AB}$. A second parameter $\xi$
has been introduced to account for the background noise in the
detection.
\begin{figure}
\includegraphics[angle=-90,width=.425\textwidth]{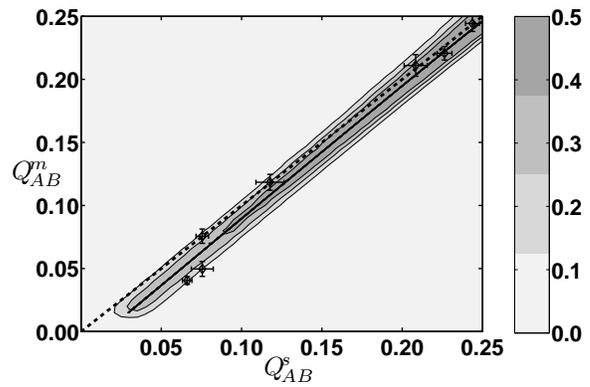}
\caption{\label{fig:QAB} Plot of $Q_{AB}^{m}$ vs $Q_{AB}^{s}$, as
defined in the text. The experimental points along with their
statistical errors are reported as diamonds. The dashed-line
represents the relation given in Eq.\eqref{Eq:QAB_Z}. The
solid-line is drawn by setting the parameters, introduced in the
text, $\Delta=0.015$ and $\xi=0.03$. Results for a uniform random
distribution of these parameters in the ranges $[0,0.03]$ and
$[0,0.06]$, respectively, are reported as a contour plot for
$N_{t} = 5\times10^{5}$ simulated trials. The contour plot
represents the frequency of the trials in the bins, equally spaced
with area $0.005\times0.005$ normalized to the maximum.}
\end{figure}
In Fig.\ref{fig:QAB} we plotted $Q^{m}_{AB}$ vs $Q^{s}_{AB}$,
i.e. the averages of the quantities respectively present on the
left and on the right side of Eq.\eqref{Eq:QAB_Z} over all the
state preparations. The dashed-line represents the linear relation
given by Eq.\eqref{Eq:QAB_Z}. The slightly sloped band is the
result of a numerical simulation with values for the  parameters $\Delta$ and
$\xi$ reported in caption of Fig.~\eqref{fig:QAB}.
%%%%%%%%%%%%%%%% Mutal Informations %%%%%%%%%%%%%%%%%%%%
\\
\indent In order to prove the security of our setup we must
compare the information shared by Alice and Bob with the one
possessed by Eve. It is known~\cite{Csiszar78} that a secret key
can be safely distilled with unidirectional classical
communication if the condition $I_{AB}\geq
\textrm{min}[I_{AE},I_{BE}]$ is accomplished. The average
Alice-Bob mutual information is given by:
\begin{equation}\label{IAB}
    \overline{I}_{AB}
        =\left(  I_{ABz}+I_{ABx}\right)/2
\end{equation}
where $I_{ABi}=1-h\left(Q_{ABi}\right)$ with $i = (x,z)$, and
$h(x) = x\,\log_2(x) + (1-x)\,\log_2(1-x)$ is the binary
entropy.
%~\cite{Gisin02}.
\begin{figure}
\includegraphics[angle=-90,width=.425\textwidth]{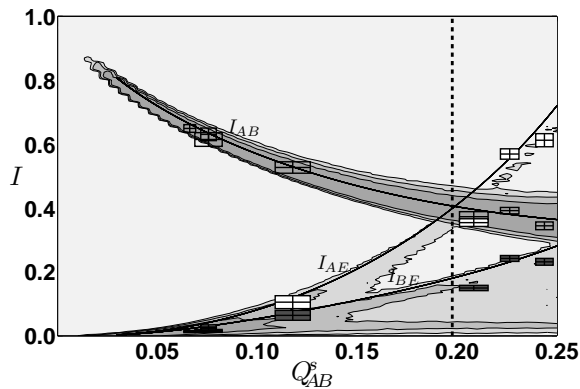}
\caption{\label{fig:I} Mutual information as function of the QBER $\rm{Q}_{AB}^{s}$~.
The experimental points are reported with their statistical errors
as crossed-rectangles (${\bar I}_{AB}$),
white-crossed-rectangles (${\bar I}_{AE}$) and
gray-crossed-rectangles (${\bar I}_{BE}$). The solid lines
are the mutual information  with $\Delta=0.015$ and
$\xi=0.03$.
Results of $N_{t} = 5\times10^{5}$ simulated trials
with the same parameters used for Fig.\ref{fig:QAB} are reported. The asymmetric imperfection of the two channels cause the gray area below the lines $I_{AE}$, $I_{BE}$.}
\end{figure}
To evaluate $I_{AE}$ we need an estimate of
$Q_{AE}^{\ZZ,\XX}$, defined as the error rate between Eve's
guesses on Alice's encoding and Alice's real encoding.
For Eve's $\ZZ$-attack, Eq.\eqref{Z}, $Q_{AE}^{\ZZ}$ reads:
 \begin{eqnarray}
  Q_{AE}^{\ZZ}
    = \frac{1}{2}-2\sqrt{q_{1x}q_{2x}\left( 1-q_{1x}\right)
        \left( 1-q_{2x}\right) }\,.
\end{eqnarray}
A similar result holds for $Q_{AE}^{\XX}$. We note that these two
quantities are independent of initial basis preparation, and
depend only on non-orthogonality of Eve's ancillae. The average
Alice-Eve mutual information is then
\begin{equation}\label{IAE}
  \overline{I}_{AE} = (I_{AE}^{\ZZ}+I_{AE}^{\XX})/2,
\end{equation}
where $I_{AE}^{\ZZ,\XX}=1-h(  Q_{AE}^{\ZZ,\XX})$. An expression
for the mutual QBER between Bob and Eve, $Q_{BE}$, can be derived
in terms of $Q_{AB}$ and $Q_{AE}$ using the relation $
Q_{BE}=Q_{AB}+Q_{AE}-2Q_{AB}\cdot Q_{AE}$~\cite{Lucamarini05}. The
average mutual information between Bob and Eve is then expressed
as:
\begin{equation}\label{IBE}
    \overline{I}_{BE}
        =(I_{BEz}^{\ZZ}+I_{BEx}^{\ZZ}+I_{BEx}^{\XX}+I_{BEz}^{\XX})/4\,,
\end{equation}
where $I_{BEi}^{(\ZZ,\XX)}=1-h(Q_{BEi}^{\ZZ,\XX})$, with
$i=(x,z)$.
\\
\indent In Fig.\ref{fig:I} are plotted the curves of mutual
information $I$ according to Eqs.(\ref{IAB}),~(\ref{IAE})
and~(\ref{IBE}) as function of the quantity $Q_{AB}^{s}$, already
defined. We report in the same figure experimental, numerical and
theoretical values. The solid lines represent our best fit of the
experimental data. It is worthy of note the almost perfect
intersection of these lines for ${\bar I}_{AB}$, ${\bar I}_{AE}$
and the theoretical (dashed) line at $Q_{AB}^{s}\simeq19\%$,
corresponding to the $\simeq23\%$ of detection probability in
Ref.~\cite{Lucamarini05}. Furthermore the curve for ${\bar
I}_{BE}$ is always below ${\bar I}_{AB}$, implying the security of
the scheme against IIA regardless of the noise on the channel.
%%%%%%%%%% Conclusions %%%%%%%%%%%
\\
\indent In conclusion we reported on the experimental test of the
two-way deterministic protocol for quantum communication LM05. We
modulated the noise on the channel in such a way as to simulate
the disturbance introduced by Eve's IIA. By means of independent
measurements of the various involved QBERs we proved the soundness
of Eq.\eqref{Eq:QAB_Z} and rated the quality of our setup. With a
subsequent measure we estimated the mutual information between
Alice, Bob and Eve. Although we did not perform a direct,
contextual, transmission of a string of bits we believe that the
good agreement between experimental data and theoretical
predictions presented in this work witnesses its potential
feasibility.\\
\indent We thank S. Mancini, D. Vitali and S. Pirandola for
fruitful discussions. This work has been supported by the
Ministero della Istruzione, dell' Universit\`a e della Ricerca
(FIRB-RBAU01L5AZ and PRIN-2005024254), and the European Commission
through the Integrated Project `Qubit Applications' (QAP),
Contract No 015848, funded by the IST directorate.

%%%%%%%%  Bibliography  %%%%%%%%%%


\begin{thebibliography}{99}
\bibitem{foot1} With `in principle' we essentially mean `on a noiseless and lossless channel'.
\end{thebibliography}
\end{document}